\newcommand{\be}{\begin{equation}}\newcommand{\ee}{\end{equation}}
\newcommand{\bea}{\begin{eqnarray}}\newcommand{\eea}{\end{eqnarray}}
\newcommand{\nn}{\nonumber}\newcommand{\p}[1]{(\ref{#1})}

\newcommand{\cD}{{\cal D}}
\newcommand{\cDb}{\bar{\cal D}}
\documentstyle[12pt]{article}

\begin{document}
\thispagestyle{empty}
\vspace{2cm}
\begin{flushright}
LPTHE 00-21\\
JINR E2-2000-134\\
ITP-UH-10/00\\
hep-th/0006017\\
May, 2000\\[3cm]
\end{flushright}
\begin{center}
{\Large\bf Double-vector multiplet and partially broken\\ N=4, d=3
supersymmetry }
\end{center}

\vspace{1cm}

\begin{center}
{\large\bf  E. Ivanov${}^{a,b}$,
S. Krivonos${}^{a}$, O. Lechtenfeld${}^{c}$ }
\end{center}

\begin{center}
${}^a$ {\it Bogoliubov  Laboratory of Theoretical Physics, JINR, \\
141980 Dubna, Moscow Region, Russia} 

\vspace{0.2cm}
${}^b$ 
{\it Laboratoire de Physique Th\'eorique et des Hautes Energies, \\
Unit\'e associ\'ee au CNRS UMR 7589, Universit\'e
Paris 7,}\\ 
{\it  2 Place Jussieu, 75251 Paris Cedex 05, France}

\vspace{0.2cm}
${}^c$ {\it Institut f\"ur Theoretische Physik, Universit\"at Hannover,} \\
{\it Appelstra\ss{}e 2, 30167 Hannover, Germany}
\end{center}

\vspace{2cm}

\begin{abstract}
\noindent We elaborate on a new $N{=}2,\, d{=}3$ supermultiplet 
(double-vector multiplet) with a non-trivial off-shell realization of the
central charge. 
Its bosonic sector comprises two abelian gauge vector fields forming an 
$SO(2)$ vector. We present a superfield formulation of this multiplet in the
central-charge extended $N{=}2,\, d{=}3$ superspace and then employ it, in the
framework of the nonlinear realizations approach, as the Goldstone one 
for the partial breaking $N{=}4\rightarrow N{=}2$ in three dimensions. 
The covariant equations of motion for the self-interacting Goldstone superfield
arise as a natural generalization of the free ones and are interpreted  as
the worldvolume supersymmetric form of the equations of motion of a $N{=}4$
D2-brane. For the vector fields we find a coupled nonlinear system of
Born-Infeld type and demonstrate its dual equivalence to the $d{=}5$ membrane.
The double-vector multiplet can be fused  with  some extra $N{=}2,\, d{=}3$ 
multiplet to form an off-shell $N{=}4,\, d{=}3$  supermultiplet. 
\end{abstract}

\newpage
\setcounter{page}{1}
\section{Introduction}
The purely geometric target space actions for extended objects
possess a lot of gauge (super)symmetries which are all realized linearly.
After proper gauge fixing the corresponding world-volume actions
describe a very special type of highly nonlinear theories.
The main property of these theories is a nonlinear realization of
broken (super)symmetries while the unbroken ones are still linearly realized.
Among all spontaneously broken symmetries  supersymmetry plays a very
special role. Nowadays, it is clear that just the nonlinearly realized broken
supersymmetries single out the Nambu-Goto-Born-Infeld type actions as the
actions compatible with the requirement of invariance under both the linearly 
as well as the nonlinearly realized supersymmetries \cite{bw}-\cite{BMZ1}.

As usual, the partial spontaneous breaking of global supersymmetry
(PBGS) implies the presence of a massless Goldstone multiplet of the residual
unbroken supersymmetry. The choice of such a multiplet is not unique
\cite{bg2}. The proper choice of the Goldstone multiplet can greatly simplify
the construction of the invariant Goldstone superfield action. Employing the
$N=1,\, d=4$ tensor multiplet  instead of chiral multiplet gives a seminal
example of such a simplification \cite{bg3,rt,gpr}. In the case of chiral
Goldstone multiplet (and, more generally, in all cases where the number of
physical scalars in the Goldstone multiplet exceeds one), a highly nonlinear
redefinition of the Goldstone scalar superfields is required to bring the 
action to the standard form (yielding the Nambu-Goto action for scalars in the
bosonic sector). In contrast, for the case of $N=1$ tensor Goldstone 
multiplet, when
one of the physical scalars is off-shell dualized into the tensor
gauge field (``notoph''), the problem of constructing the correct Goldstone
superfield action is drastically simplified because only one
physical scalar field is present off shell. The action constructed by
the methods of  refs. \cite{bik2,rt,gpr} immediately yields the
static-gauge Nambu-Goto action for the single scalar, no redefinition of the
Goldstone superfield is needed. From this
point of view, the $N=2 \rightarrow N=1$ supersymmetry breaking scheme with
the $N=1, d=4$ vector multiplet as the Goldstone one should be
simplest, because this version contains no physical scalar fields at all.
Indeed, this is the case \cite{bg2}.

In contrast to the $N=1, d=4$ vector multiplet, its dimensionally reduced
version, the $N=2,\, d=3$ vector multiplet, contains a scalar field. Yet, in
three dimensions the scalar is dual to a gauge field, suggesting the
existence of another $N=2, d=3$ supermultiplet
which contains only two vector fields in its bosonic sector. In Section 2 
we construct such a multiplet. Decomposed into
$N=1, d=3$ superfields, it comprises two $N=1$ vector multiplets, and we
naturally name it the double-vector $N=2, d=3$ multiplet. An additional
peculiarity of this multiplet is a non-trivial off-shell realization of
the central charge which appears in the closure of the two $N=1$ 
supersymmetries.
In Section 3 we develop a superfield description of the double-vector
multiplet in $N=2, d=3$ superspace with a central charge. In Section 4 we
consider the spontaneous breaking of $N=4, d=3$ supersymmetry to $N=2, d=3$ in
the framework of nonlinear realizations and choose the double-vector multiplet
as the Goldstone one accompanying this breakdown. The nonlinear generalization
of the basic constraints on the double-vector multiplet gives us
the superfield equations of motion. The latter we interpret as the worldvolume
equations of some $N=4$ D2-brane comprising two Maxwell gauge field strengths 
in its worldvolume multiplet. In Section 5 we combine the double-vector
multiplet with some extra  $N=2, d=3$  multiplet to gain an off-shell
$N=4,d=3$ supermultiplet. By its component content, it is a dimensional
reduction of the $N=2,d=4$ vector-tensor multiplet.

\section{Double-vector multiplet in $N=1$ superspace}
In this Section we perform the direct construction of the double-vector
multiplet in $N=1, d=3$ superspace.

We start with two $N=1,d=3$ vector multiplets $\psi_a^i\; (i=1,2)$,
\be
D_a \psi^{ai}  =  0\; , \quad \Rightarrow\; \left\{ \begin{array}{l}
              D^2\psi_a^i \equiv D^c D_c \psi_a^i = 
  -i\partial_{ab}\psi^{bi} \\
              \partial_{ab}D^a \psi^{bi} =0
               \end{array} \right.  , \label{bascon} 
\ee
where
\be
D_a=\frac{\partial}{\partial \theta^a}+\frac{i}{2}\theta^b
  \partial_{ab}\; \quad \left\{ D_a,D_b\right\}= i\partial_{ab}\;,
\ee
$a,b=1,2$ being the $d=3$ $sl(2,R)$ spinor indices\footnote{The
indices are raised and lowered as follows: $V^a=\varepsilon^{ab}V_b,\;
V_b = \varepsilon_{bc}V^c,\; \varepsilon_{ab}\varepsilon^{bc}=\delta_a^c$.}. 

The main question we address here is whether it is possible to realize an
additional $N=1$ supersymmetry on $\psi_a^i$ in such a way that it forms, 
together with the manifest one, an $N=2, d=3$ supersymmetry. Assuming the
second supersymmetry to be linearly realized, the corresponding 
transformation law
of  the superfields $\psi_a^i$  appears to be 
\be
\delta \psi_a^i = \varepsilon^{ij}\epsilon^b D_b \psi^j_a  \; . \label{tr-1}
\ee
One can easily check that the standard requirement of compatibility of the
transformations \p{tr-1} with the basic constraints
\p{bascon} puts our superfields on-shell, 
\be
D^aD_b\psi_a^i=0
\;\Rightarrow\quad 
\partial_{ab}\psi^{bi}=0 \; . 
\ee
To have an off-shell supermultiplet, we should modify the transformation
law \p{tr-1}. The seemingly unique possibility of such a modification is to 
admit $\theta$-dependent terms. Requiring the standard closure of the second
supersymmetry, 
\be
\left\{ S_a, S_b \right\} = P_{ab}\;, 
\ee
together with the preservation of the constraints \p{bascon} 
completely fixes the additional terms and produces
\be
\delta \psi_a^i =\varepsilon^{ij}\left( \epsilon^b D_b \psi_{a}^j + 
  i\epsilon_c\theta^c 
         \partial_{ab} \psi^{bj} \right)  \; . \label{tr1}
\ee
Due to the explicit presence of $\theta$ in \p{tr1}, the bracket of the 
manifest
and the second $N=1$ supersymmetry yields a central charge transformation
\be\label{cctr}
 \delta_z \psi_{a}^i  = i c \varepsilon^{ij}\partial_{ab} \psi^{bj}\; 
\ee
which commutes with both supersymmetries ($c$ is the corresponding
transformation parameter). Hence,  we are facing the $N=2, d=3$ supersymmetry
algebra with a central charge in \be
\left\{ Q_a, S_b \right\} = \varepsilon_{ab}Z \; .
\ee
Thus, our double-vector multiplet requires an off-shell central charge which
vanishes on-shell.

The free action for the double-vector multiplet has the very simple form
\be
S=\int d^3x d^2 \theta \left( \psi^{ai}\psi^i_a  \right) \;.
\ee
It reveals a manifest internal $SO(2)$ symmetry which commutes with 
both supersymmetries and acts on the index $i$. 
All the component fields in $\psi_a^i$,
including two Maxwell $d=3$ field strengths, are $SO(2)$ vectors.   

\section{Double-vector multiplet in $N=2$ superspace}
Since the double-vector multiplet has a non-zero central charge, it is 
necessary to enlarge the usual $N=2$ superspace 
$\left\{ x^{ab},\theta^a,\bar\theta^a \right\}$ by one extra bosonic 
coordinate $z$.
The algebra of covariant derivatives in this extended
superspace has the following form:
\be
\left\{ \cD_a , \cDb_b \right\} = 
 i\partial_{ab}- \varepsilon_{ab}\partial_z\;, \;
\left\{ \cD_a , \cD_b \right\}=0 \; , \; \left\{ \cDb_a , \cDb_b \right\}=0\; ,
\ee
where  $\cD_a, \cDb_b$ read
\bea\label{flatcd}
&& \cD_a=D_a+\frac{1}{2}{\bar\theta}_a
  \partial_z \equiv 
 \frac{\partial}{\partial \theta^a} -
\frac{i}{2}{\bar\theta}{}^b\partial_{ab}+ \frac{1}{2}{\bar\theta}_a
  \partial_z\;,\nn \\
&& \cDb_a= {\bar D}_a+\frac{1}{2}{\theta}_a
  \partial_z\equiv
-\frac{\partial}{\partial {\bar\theta}{}^a} +
\frac{i}{2}{\theta}{}^b\partial_{ab}+ \frac{1}{2}{\theta}_a
 \partial_z\;.
\eea
The supersymmetry transformations are
\be
\delta x^{ab}=\frac{i}{4}\left( \varepsilon^{(a}{\bar\theta}^{b)}+
  {\bar\varepsilon}^{(a}\theta^{b)}\right) \;, \;
\delta z = -\frac{1}{2}\left( \varepsilon^a{\bar\theta}_a-
 {\bar\varepsilon}^a\theta_a \right)\; , \;
\delta \theta^a=\varepsilon^a\;, \;
\delta\bar\theta^a = \bar\varepsilon^a \;.
\ee
The double-vector multiplet in $N=2, d=3$ central-charge extended superspace 
can be described by a chiral fermionic superfield $\eta_a$:
\be\label{vv1}
\cD_a {\bar\eta}_b =0 \; , \;  \cDb_a {\eta}_b =0
\ee
subject to the additional constraints
\be\label{vv2}
\cD^a \eta_a = 0\; , \; \cDb^a {\bar\eta}_a =0 \;.
\ee
As a consequence of \p{vv1},\p{vv2} and their corollaries 
\be \label{corol}
\cD^2\eta^a = \cDb^2 \bar\eta^a = 0
\ee
the expansion of
$\eta_a$ in the anti-commuting coordinates is ``ultra-short'',
\be\label{etacomp}
\eta_a= \nu_a+\theta^b A_{ab}+\frac{1}{2}\theta^b{\bar\theta}^c\left( 
 \varepsilon_{bc}\partial_z\nu_a-i\partial_{bc}\nu_a \right) \;,
\ee
where $A_{ab}$ is further restricted to be a Maxwell-field strength
\be
A_{ab}=A_{ba}\;, \quad \partial_{ab}A^{ab}=0 \; .
\ee

It is worth emphasizing that the constraints \p{vv1}, \p{vv2} entirely fix
the dependence of the superfields $\eta_a,{\bar\eta}_a$ on the central charge
coordinate $z$:
\be
\partial_z \eta_a = -i\partial_{ab}\eta^b \; , \quad
\partial_z \bar\eta_a = i\partial_{ab}\bar\eta^b \; \quad \rightarrow \quad 
\partial_z \nu_a = -i\partial_{ab}\nu^b~, \; \partial_z A_{ab} =
-i\partial_{ac}A^c_b . 
\ee

Thus, our complex $N=2, d=3$ superfield $\eta_a$ subject to the constraints
\p{vv1},\p{vv2} is just the same 
double-vector multiplet considered in the previous Section.

The free action for the double-vector multiplet in the central-charge extended
$N=2$ superspace can be written as 
\be\label{freevv}
S=
\int d^3x d^4 \theta\; (\theta{\bar\theta}) \left( \eta {\bar\eta} \right)\;.
\ee
Despite the presence of the explicit $\theta$ dependence in \p{freevv}, the
action is supersymmetric as a consequence of the above constraints. One
of these consequences is the property that the Lagrangian density $L_{free} =
\eta^a\bar\eta_a$ in \p{freevv} obeys the constraints    
\be \label{corol1}
\cD^2 L_{free}= \cDb^2 L_{free} = 0 \; .
\ee
We also note that in \p{freevv} one can use the integration measure
of the standard $N=2$ superspace without central charge, 
$$
d^3x d^4\theta =\frac{1}{16} d^3x (D^aD_a)({\bar D}_b{\bar D}^b)\;,
$$
because the derivative of the
Lagrangian density with respect to $z$ is a total $x$-derivative: 
\be
\partial_z L_{free} =
 i\partial_{ab} \left( \eta^a {\bar\eta}^b \right)\; .
\ee
In the components \p{etacomp} the action \p{freevv} reads
\be\label{ca}
S=\frac{1}{2}\int d^3x \left( - i\nu^a\partial_{ab}\bar\nu^b +
\frac{1}{2}A^{ab}{\bar A}_{ab} \right) \; .
\ee
Hence, the free equations of motion which follow from  \p{ca}
\be
\partial_{ab}\nu^b=0\; ,\quad \partial_{ac}A^c_b+\partial_{bc}A^c_a=0
\ee
are equivalent to discarding the central charge dependence of the 
double-vector multiplet in the basic constraints \p{vv1}, \p{vv2}.
Thus the latter together with 
\be \label{Zdisc}
\partial_z\eta_a = \partial_z\bar\eta_a = 0~,
\ee
represent the superfield form of the equations of motion.
\setcounter{equation}{0}

\section{$N=4 \rightarrow N=2 $ PBGS with double-vector multiplet}
As one of the possible applications of the double-vector multiplet, we can try
to utilize it as the Goldstone superfield describing the partial breaking
$N=4 \rightarrow N=2$ in $d=3$. Indeed, it contains two fermionic fields 
which could be identified with the Goldstone fermions of two broken
$N=1, d=3$ supersymmetries. Then $\eta_a, \bar\eta_a$ can be treated as the
superfield Goldstone fermions associated with the spontaneously broken
supersymmetry generators. 

Leaving aside, for the time being, the search for the
corresponding Goldstone off-shell superfield action generalizing \p{freevv}, we
pursue here a different goal. Namely, we shall get the
corresponding superfield equations of motion  (together with the Bianchi
identities) starting from the nonlinear realization of
the global supersymmetry group along the lines of \cite{bik1}. As far as we are
interested in the equations of motion, we can discard the central charge
dependences of our superfields. Hence, our starting point will be the
 $N=4, d=3$
Poincar\'{e} superalgebra {\it without} central charges: 
\be\label{n4sp}
\left\{ Q_a, {\bar Q}_b\right\}= -P_{ab}\; ,\quad \left\{ S_a, {\bar
S}_b\right\}= -P_{ab}\; . 
\ee 
Assuming the $S_a, {\bar S}_a $ supersymmetries to
be spontaneously broken, we introduce the Goldstone superfields $\xi^a(x,
\theta,\bar\theta), \bar\xi^a(x,\theta,\bar\theta)$ as the parameters of the
following coset 
\be\label{n4coset}
g=e^{ix^{ab}P_{ab}}e^{i\theta^aQ_a-i\bar\theta^a {\bar Q}_a} 
e^{i\xi^aS_a-i\bar\xi^a {\bar S}_a} \; . 
\ee 
Using the Cartan forms
\bea
&& g^{-1}dg = i\omega^{ab} P_{ab}+i\omega_Q^a Q_a -i{\bar\omega}_Q^a{\bar Q}_a
+i\omega_S^a S_a -i{\bar\omega}_S^a{\bar S}_a \;, \nn \\
&&\omega^{ab}= dx^{ab}+\frac{i}{4}\left(
d\theta^{(a}\bar\theta^{b)}+d\bar\theta^{(a}\theta^{b)}+
d\xi^{(a}\bar\xi^{b)}+d\bar\xi^{(a}\xi^{b)} \right)\; , \nn \\
&&\omega_Q^a=d\theta^a\;,\; \bar\omega_Q^a=d\bar\theta^a \; , \;
\omega_S^a=d\xi^a\;,\; \bar\omega_S^a=d\bar\xi^a \; , 
\eea
one can define the covariant derivatives
\bea
\nabla_{ab}&=& \left( E^{-1}\right)_{ab}^{cd}\partial_{cd} \; , \nn \\
\nabla_a & = & D_a -\frac{i}{2}\left( \xi^b D_a\bar\xi^c+
  \bar\xi^b D_a \xi^c \right) \nabla_{bc} \; , \nn \\
\bar\nabla_a & = & {\bar D}_a -
  \frac{i}{2}\left( \xi^b{\bar D}_a\bar\xi^c+
  \bar\xi^b {\bar D}_a \xi^c \right) \nabla_{bc} \; , \label{n4cd} 
\eea
where 
\be
E_{ab}^{cd}=\frac{1}{2}\left( \delta_a^c\delta_b^d+\delta_a^d\delta_b^c
-\frac{i}{2}\xi^{(c} \partial_{ab}\bar\xi^{d)}
-\frac{i}{2}\bar\xi^{(c} \partial_{ab}\xi^{d)} \right)
\ee
and $D_a, {\bar D}_a$ are ``flat'' covariant derivatives \p{flatcd}
without central charge.
The covariant derivatives obey the following algebra
\bea\label{algebracd}
&& \left\{ \nabla_a , \bar\nabla_b \right\} =i\nabla_{ab} -i \left(
 \nabla_a\xi^m\bar\nabla_b\bar\xi^n +\nabla_a\bar\xi^m\bar\nabla_b\xi^n
\right)\nabla_{mb} \; , \nn \\
&& \left\{ \nabla_a , \nabla_b \right\} = -i \left(
 \nabla_a\xi^m\nabla_b\bar\xi^n +\nabla_a\bar\xi^m\nabla_b\xi^n
\right)\nabla_{mb} \; , \nn \\
&& \left\{ \bar\nabla_a , \bar\nabla_b \right\} = -i \left(
 \bar\nabla_a\xi^m\bar\nabla_b\bar\xi^n +
 \bar\nabla_a\bar\xi^m\bar\nabla_b\xi^n
\right)\nabla_{mb} \; .
\eea
Now we are ready to place additional, manifestly covariant constraints on
our Goldstone superfields $\xi_a,\bar\xi_a$. Following \cite{bik1}, we should
covariantize the ``flat'' equations of motion. We have seen that the basic
constraints  for the double-vector multiplet \p{vv1}, \p{vv2} augmented
with the condition of indepenedence on the central charge coordinate,
eq.\p{Zdisc}, give us the on-shell multiplet. 
Thus the sought-for covariantized system is 
\be\label{vvcov} \nabla_a {\bar\xi}_b =0 \; , \;  \bar\nabla_a
{\xi}_b =0\; , \; \nabla^a \xi_a = 0\; , \; \bar\nabla^a {\bar\xi}_a =0 \;. \ee
As it usually happens in the nonlinear realizations approach, 
the equations \p{vvcov} are, despite their 
simple form, rather complicated, due to
the highly nonlinear structure of the covariant derivatives. The complete
analysis of these equations is beyond the scope of the present paper, but 
in order to
understand which kind of dynamics they encode it is instructive to consider
their bosonic limit. We find that they amount to the following equations
for the vectors
$
F^{ab}\equiv \nabla^a\xi^b|_{\theta=0} ,  
 {\bar F}^{ab}\equiv -\bar\nabla^a\bar\xi^a|_{\theta=0}
$:
\be\label{n4bi}
\partial_{ab}F^b_c+F_b^m{\bar F}_a^n\partial_{mn}F^b_c=0 \; , \quad
\partial_{ab}{\bar F}^b_c+F_a^m{\bar F}_b^n\partial_{mn}{\bar F}^b_c=0 \; .
\ee
Like the previously treated cases \cite{bik1}, the system \p{n4bi} is
a disguised form of the Born-Infeld equations for two Maxwell gauge field
strengths,  augmented with the
Bianchi identities. Indeed, after rewriting eqs.\p{n4bi} as
\bea\label{n4biA}
&&\left( 1-\frac{1}{4}F^2{\bar F}^2\right)\partial_{ab}F^b_c-
  \frac{1}{2}{\bar F_a^c}
\partial_{bc}\left( F^2\right)-\frac{1}{4}\left( {\bar F}^2\right)
  F_b^c\partial_{ac}\left( F^2\right)=0 \; , \nn \\
&&\left( 1-\frac{1}{4}F^2{\bar F}^2\right)\partial_{ab}{\bar F}^b_c-
 \frac{1}{2} F_a^c
\partial_{bc}\left( {\bar F}^2\right)-\frac{1}{4}\left(  F^2\right)
  {\bar F}_b^c\partial_{ac}\left( {\bar F}^2\right)=0 \; , 
\eea
one can bring them into the following equivalent form
\bea
&&\partial_{ab}V^{ab}=0\; ,\;\partial_{ab}{\bar V}^{ab}=0\; ,\label{bianchi}\\
&&\partial_{ab}G^b_{c}+ \partial_{cb}G^b_{a}=0 \; , \; 
\partial_{ab}{\bar G}^b_{c}+\partial_{cb}{\bar G}^b_{a}=0 \; , \label{n4bieom}
\eea
where
\bea
&& V^{ab}=\frac{4F^{ab}+2F^2{\bar F}^{ab} }{4- F^2{\bar F}^2}\;,\; 
{\bar V}^{ab}=\frac{4{\bar F}^{ab}+2{\bar F}^2 F^{ab} }
    {4- F^2{\bar F}^2}\;,\nn \\
&&G^{ab}=\frac{4F^{ab}-2F^2{\bar F}^{ab} }{4- F^2{\bar F}^2}\;,\; 
{\bar G}^{ab}=\frac{4{\bar F}^{ab}-2{\bar F}^2 F^{ab} }
    {4- F^2{\bar F}^2}\;.
\eea
After introducing the ``genuine'' field strengths $V^{ab}, {\bar V}^{ab}$
eqs. \p{bianchi} are recognized as the Bianchi identities, while
eqs. \p{n4bieom} acquire the familiar form with
\be
G^{ab}=\frac{ \left( 1+ V{\bar V}\right) V^{ab}-V^2 {\bar V}^{ab} }
{\sqrt{ \left( 1 + V{\bar V}\right)^2-V^2{\bar V}^2 }}\;, \;
{\bar G}^{ab}=\frac{ \left( 1 + V{\bar V}\right){\bar V}^{ab}-
 {\bar V}^2  V^{ab} }
{\sqrt{ \left(1 + V{\bar V}\right)^2-V^2{\bar V}^2 }}\; .
\ee
Thus, in this new basis the action for the bosonic core is the Born-Infeld
type action for the $SO(2)$ invariant system of two $d=3$ gauge field
strengths $V, \bar V$ \footnote{We use the $d=3$ Minkowski signature 
$(+ - -)$ and normalize the vectors so that $A_mB^m = A_{ab}B^{ab}$.}  
\bea 
S &=& \int d^3 x \left\{ \sqrt{ \left(1 + V{\bar V}\right)^2-V^2{\bar V}^2 } -
1 \right\} \nonumber \\ 
&=& \int d^3 x \left\{ \sqrt{\mbox{det}\left[
\eta_{mn} + (V_m \bar V_n + \bar V_m V_n) \right]} - 1\right\} \; .
\label{so2BIact} \eea 
(up to an overall coupling constant). Note that in the limit $V_m = \bar V_m$
it does not immediately yield the standard Born-Infeld action with 
Lagrangian density 
$$\sim \sqrt{\mbox{det}\left[\eta_{mn} + \sqrt{2}\epsilon_{mnk}V^k
\right]},
$$ 
but it is easy to show that these two reperesentations are equivalent.  

Since eqs. \p{vvcov} represent an $N=4$
supersymmetrization of the Born-Infeld equations \p{bianchi}, \p{n4bieom}, 
they can be interpreted from the brane's viewpoint
as manifestly worldvolume supersymmetric equations 
for some space-time filling $N=4$ 
supersymmetric D2-brane. This brane possesses an internal
rigid $SO(2)$ symmetry and incorporates two abelian gauge fields in its
worldvolume multiplet, such that they form an $SO(2)$ vector. Note that this
$SO(2)$ naturally arises in the nonlinear realization setting as the
automorphism group of the spontaneously broken part of $N=4, d=3$ 
supersymmetry realized on the generators $S_a, \bar S_b$ in \p{n4sp}. It is
easy to show (see below) that the action \p{so2BIact} is dual to the
static-gauge form of the Nambu-Goto action for the membrane in $D=5$, with an
$SO(2)$ doublet of scalar fields.  So it is plausible that the full superfield
system is dual to the $N=1, D=5$ scalar supermembrane which in the static gauge
has just two transverse bosonic directions. In this picture, the above $SO(2)$
reappears as the rotation group of the transverse  directions, a remnant of
the full  $D=5$ Lorentz group $SO(1,4)$ corresponding to the breakdown pattern
 $SO(1,4) \rightarrow SO(1,2)\times SO(2)$. This supermembrane can be also
obtained via the double dimensional reduction from the $N=1, D=6$ scalar
3-brane which, from the $d=4$ world-volume point of view, 
is the theory of partial breaking $N=2
\rightarrow N=1$ with  a chiral Goldstone superfield \cite{bg1}. Note that
the ``genuine'' $N=4$ D2-brane built with the worldvolume $N=2, d=3$ vector
multiplet seems not to  admit a straightforward off-shell dualization to the
considered system.  Indeed, this D2-brane clearly lacks the internal off-shell 
$SO(2)$ symmetry which is manifest in the system described here. The duality
equivalence is expected to be restored on-shell, with the $SO(2)$ becoming 
a kind of duality rotation. Of course, all these statements can be  made more
precise after constructing an action for the above system and  rederiving
eqs. \p{vvcov} just from it.  

As the last topic of this Section let us sketch the proof of the duality
equivalence of the BI action \p{so2BIact} to the static-gauge Nambu-Goto action
for two  scalar bosons. 

One adds the Bianchi identities \p{bianchi} with Lagrange multipliers $\phi,
\bar\phi$ to \p{so2BIact}, passing to the equivalent action 
\be \label{BIactD}
\tilde S = \int d^3 x \left\{\sqrt{ \left(1 + V{\bar V}\right)^2-V^2{\bar V}^2
} -1 + \partial\phi \bar V + \partial\bar\phi V \right\}\; . 
\ee 
Then one varies with respect to $\bar V^{ab}, V^{ab}$ which are
unconstrained in \p{BIactD} and gets for them the following algebraic equation
\be \label{algeq}
\frac{1}{\sqrt{ \left(1 + V{\bar V}\right)^2-V^2{\bar V}^2 }}\left[ \left(1+
V\bar V \right)V_{ab} - V^2 \bar V_{ab} \right] +\partial_{ab}\phi = 0 
\ee
and its conjugate. These equations can be solved for $V, \bar V$, expressing
the latter in terms of $\partial_{ab}\phi, \partial_{ab}\bar\phi$. This is
rather straightforward. The simplifying observation is that these equations 
immediately imply 
$$
V\partial\bar\phi = \bar V \partial \phi~ 
$$
which, after substituting into \p{BIactD}, allows one to represent the dual
action  in terms of the two invariants $V\bar V$ and $V^2\bar V^2$, namely
\be \label{BIactD1}
\tilde S = \int d^3 x \left\{ \frac{1 + V^2\bar V^2- (V\bar V)^2}{\sqrt{
\left(1 + V{\bar V}\right)^2-V^2{\bar V}^2 }}-1 \right\}~.
\ee 
Thus, to get the dual action of $\phi, \bar \phi$, it is sufficient to find
the expressions for these invariants.  They are obtained from \p{algeq} in a
rather simple form which we, however, omit to present here. After substituting
these expressions into \p{BIactD1}, one arrives at the action  
\be
\label{BIactD1a} \tilde S = \int d^3 x \left\{ \sqrt{ \left(1 - \partial\phi
\partial \bar\phi \right)^2- (\partial \phi)^2 (\partial \bar\phi)^2} - 1
\right\}  \ee 
which is recognized as the static-gauge form of the standard Nambu-Goto 
action for the membrane in 5-dimensional Minkowski space: 
\be \label{NG}
\tilde S = \int d^3 x \left\{ \sqrt{\mbox{det}\left[ \eta_{mn} - (\partial_m
\phi \partial_n\bar\phi + \partial_m\bar\phi\partial_n\phi )
\right]} - 1\right\}~.    
\ee

\setcounter{equation}{0}
\section{A new $N=4, d=3$ supermultiplet}
The double-vector multiplet we considered in the previous Sections is very
similar to the $N=2, d=4$ vector-tensor multiplet \cite{ssw}. It is 
known \cite{ssw} that the vector-tensor multiplet can be combined with
Fayet-Sohnius hypermultiplet to form an off-shell $N=4$ SYM supermultiplet
in the central charge extended superspace. Therefore, a natural question
arises in our case: is it  possible to add some additional
$N=2, d=3$ supermultiplet to our double-vector multiplet so as to end up 
with an $N=4,d=3$ off-shell supermultiplet?

Some hints how to construct such $N=4$ supermultiplet come from the
previous Section. Namely, we showed that the double-vector
multiplet can be utilized as the Goldstone superfield for a nonlinear
realization of $N=4, d=3$ supersymmetry partially broken down to $N=2$. This
gives us the theory of a self-interacting double-vector multiplet, whose
Lagrangian density,  a nonlinear generalization of \p{freevv}, 
could be in principle constructed in terms of $\xi^a,
\bar\xi^a $. We  also know that the PBGS theories often allow one
to combine the Lagrangian density together with the basic Goldstone superfields
into a linear  supermultiplet of higher supersymmetry. These facts suggest
a combination of our double-vector supermultiplet 
with some scalar bosonic superfield $\Phi$ to form  a linear $N=4$ multiplet. 
Of course, this
additional superfield  $\Phi$ should  also be properly constrained. 

Let us now try to find a linear realization of an extra $N=2, d=3$
supersymmetry on the double-vector multiplet $\eta^a,{\bar\eta}^a$ with
\p{vv1}, \p{vv2} together with some, yet unconstrained, scalar superfield
$\Phi(x,z,\theta,\bar\theta )$. Assuming the second supersymmetry to be
spontaneously broken and requiring its standard closure, one can
find the following realization
\be\label{linn4}
\delta\eta_a=\epsilon_a-\epsilon^b\cDb_b\cD_a \Phi\; , \;
\delta\bar\eta_a=\bar\epsilon_a-\bar\epsilon^b\cD_b\cDb_a \Phi\; , \;
\delta \Phi = \epsilon^a\bar\eta_a+\bar\epsilon^a\eta_a \; .
\ee
Now we have to put the proper constraints on the superfield $\Phi$.

First, if we wish to consider $\Phi$ as a candidate for
the off-shell Lagrangian density corresponding to the PBGS system of the
previous Section, we should impose the constraints 
\be\label{n4lincon1}
\cD^2\Phi=\cDb^2\Phi=0 \ee
(recall \p{corol1}). The constraints  \p{n4lincon1} are in agreement with the
transformation properties of $\Phi$ \p{linn4}.

The rest of constraints follow from the condition that the r.h.s. of
$\delta\eta^a, \delta\bar\eta^a$ in \p{linn4} obeys the same constraints
as $\eta^a,\bar\eta^a$ in  eqs.\p{vv1}, \p{vv2}. This implies
\be\label{n4lincon2}
\left(i\partial_{ab}+\varepsilon_{ab}\partial_z\right)\cD^b\Phi=0\; ,\quad
\left(i\partial_{ab}-\varepsilon_{ab}\partial_z\right)\cDb^b\Phi=0\; .
\ee
One may check that these constraints are also compatible with
the transformation properties of $\Phi$ \p{linn4}. Thus, the $N=2$
superfields $\left\{ \eta^a,\bar\eta^a, \Phi \right\}$, defined in the
central charge extended superspace and respecting the constraints
\p{vv1}, \p{vv2}, \p{n4lincon1}, \p{n4lincon2}, constitute indeed a linear 
off-shell $N=4$ supermultiplet. One $N=2$ supersymmetry is manifest, while
another one is realized according to \p{linn4}.

It is instructive to consider the component structure of $\Phi$. 
Straightforward computation shows that the $\theta, \bar\theta$
expansion of $\Phi$ is also extremely short:
\be\label{Phi}
\Phi=\phi+\theta^a\bar\psi_a+\bar\theta^c\psi_a+\theta^a\bar\theta_a\rho
+\theta^a\bar\theta^b\partial_{ab}v \;,
\ee
where the $z$-dependence of the components is restricted by
\bea
&&i\partial_{ab}\bar\psi^b+\partial_z \bar\psi_a=0 \; , \;
i\partial_{ab}\psi^b-\partial_z\psi_a =0 \; , \nn \\
&& \Box \phi=2\partial^2_z\phi \; , \;
\partial_z \rho=\frac{i}{2} \Box v\; , \;
\partial_z v =-i\rho \;.
\eea
Thus, $\Phi$ contains four scalars 
$\left\{ \phi,\partial_z\phi,\rho,v\right\}$ and four fermions
$\left\{ \psi_a,\bar\psi_a\right\}$. This content coincides with that of 
$N=2, d=4$ vector-tensor multiplet \cite{ssw}, suggesting that the
supermultiplet $\left\{ \eta^a, {\bar\eta}^a,\Phi \right\}$ could be
re-obtained from the vector-tensor one via  $d=4
\rightarrow d=3$ dimensional reduction. 

Finally, we write the $N=4$ invariant free action for our multiplet as
\be \label{N4action}
S=\int d^3xd^4\theta \left( \theta\bar\theta\right)
\left( \eta^a\bar\eta_a + \cD^a \Phi \cDb_a\Phi \right) \; .
\ee
Like in the case of the action \p{freevv}, the invariance of \p{N4action} 
is checked with the heavy use of the basic constraints on the 
superfields involved.  

\setcounter{equation}0
\section{Concluding remarks}
In this paper we have constructed, for the first time, a new off-shell 
$N=2, d=3$ supermultiplet which contains only vector fields
among its bosonic components. 
This ``double-vector'' multiplet possesses a non-trivial
realization of the central charge and, therefore, can be formulated only
in the central-charge extended superspace. This multiplet fills a last gap
in the variety of all possible dualizations of the $N=2, d=3$ chiral
supermultiplet and corresponds to the dualization of both  
scalar fields into 
vectors. Using the double-vector multiplet as the Goldstone multiplet for 
$N=4\rightarrow N=2$ partial breaking, we were able to derive
the superfield equations of motion. In the bosonic sector they
yield the Born-Infeld interaction of two vector fields forming an $SO(2)$
doublet. Finally, in the $N=2, d=3$ central-charge extended superspace 
we discover a 
scalar multiplet which, together with the double-vector multiplet, forms
an off-shell $N=4$ supermultiplet.

One may ask why we should consider the rather complicated double-vector
multiplet? After all,
the $N=4\rightarrow N=2$ partial breaking of supersymmetry
in $d=3$ can be perfectly described by chiral as well as by vector multiplets
(the straightforward dimensional reduction of the results of \cite{bg2},
\cite{bg3}), while the dualization of the scalar in the vector
$N=2, d=3$ supermultiplet corresponds to the reduction from an on-shell $N=1,
d=4$ supermultiplet which cannot even be coupled to supergravity \cite{tn}. 
However, the
main goal of our investigation was to show that the supermultiplets in the
central charge extended superspace can equally be used to describe the partial
breaking of supersymmetry. Moreover, as follows from the consideration in Sect
4, one may introduce self-interactions of such multiplets not
only through a non-linear deformation of the basic constraints, like in the 
case of the VT supermultiplet in $D=4$ \cite{nlVT}, but also through their
``covariantization'' with respect to nonlinear realizations of some
higher-order supersymmetries. It is this possibility to which we would like 
to draw attention. One can hope to utilize other known (and still unknown) 
off-shell linear supermultiplets with non-trivially realized central charges 
for describing PBGS patterns of extended  supersymmetries. For instance, given
$N=4,\, d=4$ supersymmetry, they provide the only presently known off-shell
realizations of it, and one can expect \cite{ikk} that they will be suitable 
for the construction of the appropriate PBGS actions along the lines of refs.
\cite{bg1}-\cite{bik3}.  

Let us finally note that in this paper we did not attempt to construct
the full action for the $N=4, d=3$ D2-brane in the nonlinear or linear
realization approaches. This problem will be addressed elsewhere.

\section*{Acknowledgements}
We would like to thank S. Kuzenko, D. Sorokin and B. Zupnik for useful 
discussions. The work of E.I. and S.K. was supported in part by the 
grants RFBR-CNRS 98-02-22034, RFBR 99-02-18417, RFBR-DFG 98-02-001-80 and 
NATO Grant PST.CLG 974874. E.I. thanks LPTHE, Universit\'e Paris 7, for 
hospitality. S.K. is grateful to the Institute of Theoretical Physics in
Hannover for hospitality during the final stage of this work.

\end{document}